# Laser stimulated second and third harmonic optical effects in F: $SnO_2$ nanostructures grown via chemical synthetic route

Anusha[a], B. Sudarshan Acharya[a], Albin Antony[a], Aninamol Ani[a], I.V. Kityk[b],*, J. Jedryka[b], P. Rakus[b], A. Wojciechowski[b], P. Poornesh[a,*,] Suresh D. Kulkarni[c] a Department of Physics, Manipal Institute of Technology, Manipal Academy of Higher Education, Manipal, Karnataka 576104, India
b Institute of Optoelectronics and Measuring Systems, Faculty of Electrical Engineering, Czestochowa University of Technology, Armii Krajowej 17, PL-42-201 Czestochowa, Poland
c Department of Atomic and Molecular Physics, Manipal Academy of Higher Education, Manipal, Karnataka 576104, India

ABSTRACT

Laser stimulated second and third harmonic generation effects in Fluorine doped tin oxide (F:$SnO_2$) nanostructures versus the fluorine content is presented. The F:$SnO_2$ nanostructures have been fabricated at various fluorine doping concentrations by spray pyrolysis technique. The films exhibit polycrystalline nature with a preferential growth orientation along (1 1 0) diffraction plane as evident from x-ray diffraction studies. The optical transmittance of the F:$SnO_2$ films has increased from 68% to 80%. Photoluminescence studies revealed that strong violet emission peak corresponds to ~400 nm and relatively weak red emission peak at about ~675 nm was observed for all the F:$SnO_2$ films. Increase in the $\beta_{eff}$ value upon fluorine incorporation supports the applicability of the deposited films in passive optical limiting applications. The principal origin of second harmonic generation signals (SHG) for this type of nanostructures is played by the space charge density acen-tricity due to the F doping. The enhanced second and third harmonic generation signals observed on F:$SnO_2$ nanostructures endorses the credibility of these materials in various nonlinear optical trigger device applications.

## 1. Introduction

In recent days, there are wide variety applications of transparent conducting oxides that involve optical coatings for lenses or mirrors, sensors, military weapon systems, memories in computers, microelectronic and hybrid circuits etc. [1,2]. Transparent Conducting Oxide (TCO) materials like tin oxide, zinc oxide, zirconium oxide, nickel oxide, Indium oxide etc. play a prominent roledue to their high optical transparency and good electrical properties. The broad applications of TCO materials include optoelectronic devices, Low-E windows, Liquid Crystal Displays (LCDs) and plasma screen displays [3]. The excellent combination of transparency and conductivity in the materials is because of the presence of native oxygen vacancy defects which further increased upon doping with suitable dopant atoms. $SnO_2$ is an n-type TCO material and is mostly preferred due to its wide bandgap (> 3 eV), non-toxicity, chemical stability and mechanically strong nature [2,4]. When these films are doped with suitable dopants such as Fluorine, Indium, Antimony, they canbeused as transparent electrode materials in solar cells, LED, flat panel displays, electrochromic devices (ECDs) and also a possible candidate for nonlinear optical applications [4,5].

In the present work, we are synthesizing tin oxide which is a highly degenerate n-type semiconductor. This degeneracy can be achieved by the substitution of $O^{2-}$ by Fluorine atom ($F^-$) which acts as an electron donor. Choice of fluorine as a dopant is mainly on the account of the similarity in ionic radius which is of 0.133 nmcompared to oxygen that of 0.132 nm [5], hence fluorine can substitute oxygen effortlessly and results in furthermore simplicity in the doping process. F:SnO2 can be a promising substitution for Indium tin oxide (ITO) as a transparent conducting layer in solar cells and other thin film applications. It is a fact that ITO is the most widely used TCO material among all. Because of wide utilization and high production cost for ITO material, an al-ternative to this material is always an inevitable requirement. F:SnO2 could be the best TCO material in replacing ITO due to its non-stoi-chiometry, chemically inert, oxidizing and reducing properties [6]. There are several deposition techniques for F:SnO2 thin films that have been reported which includes DC and RF sputtering, molecular beam epitaxy, thermal evaporation, sol-gel and spray pyrolysis [6–8]. Among all, spray pyrolysis is an extensively used technique to prepare thin films of the desired thickness. It is a simple method with low equipment cost and easiness of adding dopants that can be employed to get the thin films of good quality and uniform thickness. There are many works available on physical properties of F:SnO2 thin films by various de-position techniques. Comparatively, less works accessible regarding the combination of physical, linear and nonlinear optical properties of F:SnO$_2$ thin films by spray pyrolysis. In this context, we have made an attempt to study the influence of fluorine doping on structural, optical, morphological and nonlinear properties of SnO$_2$ thin films. Further-more, particular interest in novel nonlinear optical materials is emer-ging recently. For this reason, we will explore the laser stimulated and second and third harmonic generation features of F:SnO$_2$ nanos-tructures for the first time.

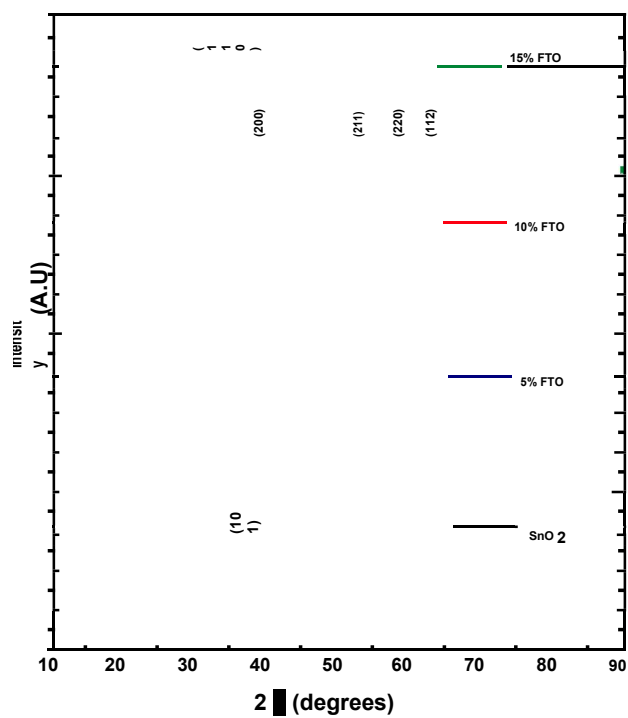

Fig. 1. XRD pattern of F:SnO$_2$ thin films.

## 2. Experimental details

For the synthesis of F:SnO$_2$ thin films, tinchloride dihydrate (SnCl$_2$·2H$_2$O) were dissolved in double distilled water to get the homogeneous solution. Ammonium fluoride (NH$_4$F) was added to the precursor solution at a ratio of 5%, 10% and 15% weight percentage in order to achieve the fluorine doping. Deposition parameters employed to prepare the film were described elsewhere [9].

The film thickness was measured using optical profilometer which is found to be around 300 nm. Structural and morphological detailswere studied by Rigaku MiniFlex600 X-ray diffractometer (Cu Kα, λ = 1.54 Å) and atomic force microscope (AFM) in tapping mode configuration. Shimadzu 1800 UV–Visible spectrometer were used to study the optical properties of F:SnO$_2$ films. Room temperature Photoluminescence (RTPL) measurements of F:SnO$_2$ films were re-corded at an excitation wavelength of 325 nm using JASCO FP-8300 spectrophotometer. The nonlinear absorption coefficient of the films was measured by implementing an open aperture Z-scan technique. It is a high sensitivity measurement technique based on spatial beam dis-tortionwhich measures absorptive and refractive nonlinearity of the films. He-Ne laser (Thorlabs HRP350-EC-1) was used as an excitation source at an input power of 19 mW and wavelength (λ) 633 nm for validating third-order nonlinear absorption properties of F:SnO$_2$ thin films. Laser-induced second and third harmonic generation measure-ments were taken before and after the laser treatment process. As a source of fundamental radiation an 8 ns Nd: YAG pulsed laser with a wavelength of 1064 nm and frequency recurrence of ~10 Hz were used. Glan's polarizer was used to tune the power of the fundamental laser. Silicon detector detects the value of the fundamental laser signal whereas Hamamatsu photomultiplier gives its second/third harmonic value based on an interferometer filter which transmits electromagnetic radiation at wavelength 355 nm (for THG) and 532 nm (for SHG).

## 3. Results and discussion

### 3.1. Structural properties

Fig. 1 represents the XRD spectra of F:SnO$_2$ films deposited at dif-ferent F doping concentrations (0, 5, 10 and 15 wt%). The different peaks on the XRD pattern indicate pure and F doped SnO$_2$ films possess polycrystalline orthorhombic crystal structure with a space group of Cmc21 (36) according to JCPDS card No. 00-013-0111 which is non-centrosymmetric [10]. It was observed that the preferential growth orientation of all the films is along (1 1 0) plane, which is in good agreement with the previously reported studies [8]. Peak intensity of the (1 1 0) plane was enhanced on F doping. Along with the major peak, other secondary peaks were observed in (2 0 0), (2 1 1), (2 2 0) and (1 1 2) planes for pristine as well as doped samples. For undoped sample, intensity reflections along (1 1 0) and (1 0 1) planes found to be strong. It is noteworthy that the crystalline nature of the SnO$_2$ films was enhanced as F incorporates into SnO$_2$ lattice. The crystallinity of the SnO$_2$ films strongly be determined by the concentration of the F in the structure and amount of F has a divergent effect on the structural properties of the SnO$_2$ films [11]. Variations on structural parameters were studied and depicted in Table 1.

It was observed that crystallite size for (1 1 0) peak has a progressive increment from 12.42 nm to 32.39 nm (for 0–15 wt% F conc.). H. Kim et al. reported a similar trend which shows crystallite size has increased from 24 nm to 30 nm (for 0 to 30 wt% F conc.) [13]. The quality of the films and defect structure of the material is defined by dislocation density [9]. In the present investigation, dislocation density is de-creasing from undoped to 15 wt% F doped SnO$_2$ films which shows that the F incorporation in the SnO$_2$ decreasing the defects of the films. A similar trend was observed for lattice strain as well as validating the reduction in the defect states.

Table 1
Structural parameters of F:SnO$_2$ thin films.

| F (wt%) | (1 1 0), 2θ | FWHM β (rad) | Crystalline Size D (nm) | Dislocation density δ (×10$^{15}$ m$^{-2}$) | Strain ε (×10$^{-3}$) |
|---|---|---|---|---|---|
| 0 | 27.31 | 0.65786 | 12.42 | 6.48 | 2.79 |
| 5 | 27.42 | 0.32532 | 25.13 | 1.58 | 1.38 |
| 10 | 27.42 | 0.26500 | 30.85 | 1.05 | 1.12 |
| 15 | 27.47 | 0.25242 | 32.39 | 0.95 | 1.07 |

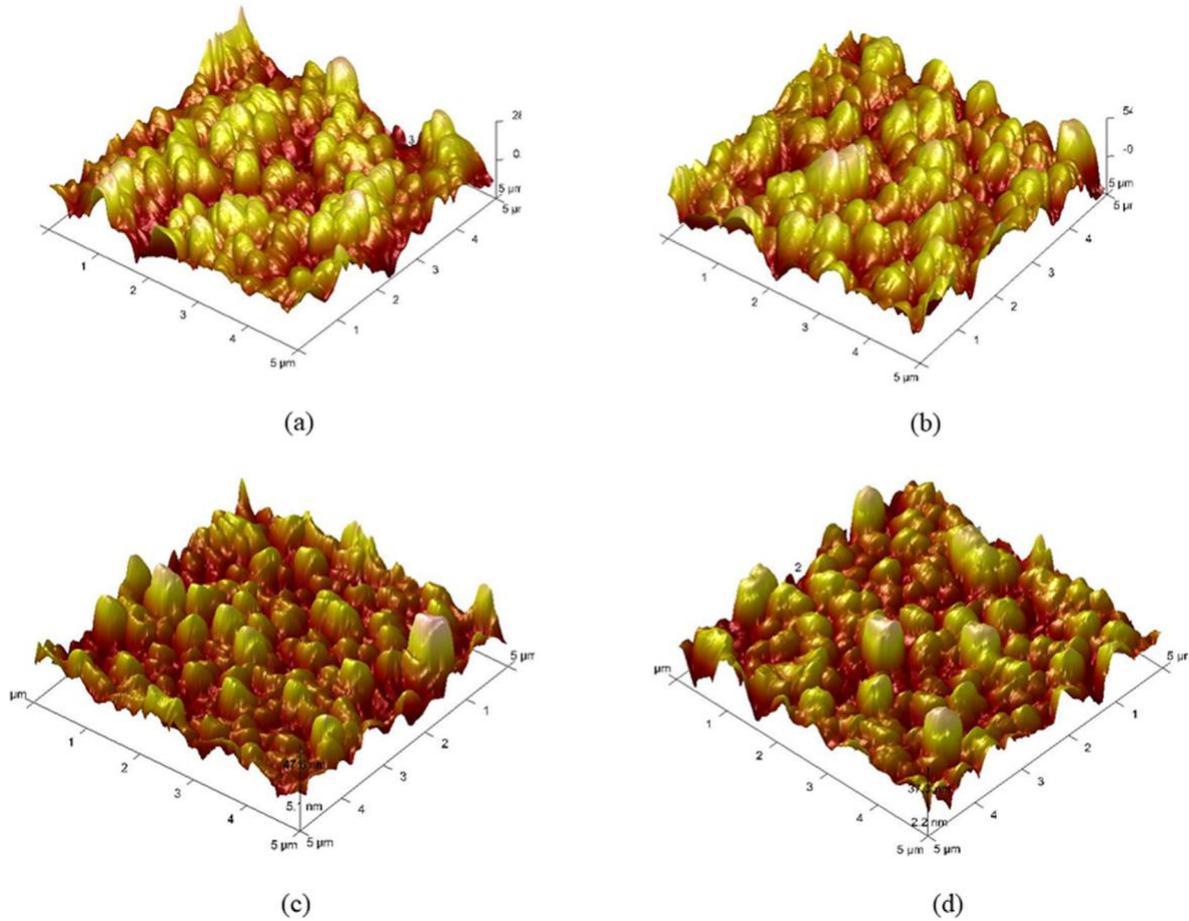

Fig. 2. AFM images of F:SnO$_2$ films at a different F doping concentration.

### 3.2. Morphological studies

AFM was used to study the surface morphology of the deposited films. Using nanoscope analysis software AFM images of F doped SnO$_2$ thin films at various doping levels were recorded in a 5 × 5 μm scan area and shown in Fig. 2. It was seen that the grain size of the SnO$_2$ thin films was increased on F doping. The reason behind the observed grain growth could be nucleation and growth rate which thus relies upon the dopant concentration and deposition parameters, that ultimately de-cides the crystallite size [14]. Variations on average surface roughness of SnO$_2$ films upon fluorine incorporations were tabulated in Table 2. The lowest roughness of 11.2 nm was obtained for 15 wt%F con-centration that exhibits a very smooth surface. AFM results confirm the XRD values whichpresent the better crystallinity of the tin oxide sam-ples on fluorine doping.

### 3.3. Linear optical properties

The transmission spectra of F:SnO$_2$ thin films were recorded at various doping concentrations (0–15 wt%) as depicted in Fig. 3(a). Absorption edges moved towards lower wavelength region as the doping level expanded from 0 to 15 wt%. Similar variation in the absorption spectra is reported by M. Thirumoorthi et al. [12]. The films exhibit the transmittance of 68–80% in the visible range of the spectrum. Enhancement in the optical transmittance is attributed to the well-crystallized film and feeble scattering from the film surface [5,14,15]. In the present case, 15 wt% F:SnO$_2$ has the highest crystallinity and the lowest value for surface roughness. Interference pattern of the films is related to the surface roughness [16], which can be seen for higher dopant amounts that possess pin-hole free and smooth surface.

The plot between $(\alpha h\nu)^2$ and photon energy (in eV) for direct allowed transitions is shown in Fig. 3(b). Tauc relation [17] gives the optical band gap of as-depositedF:SnO2 films. Table 3 provides the increased tendency of band gap values for 0–15 wt%F amount, that could be explained by Burstein-Moss shift [18,19]. i.e. increase in the fluorine concentration will cause the upward shift of Fermi level inside the conduction band, filling of lower levels by free carriers results in broadening of energy band gap [20,21].

### 3.4. Photoluminescence studies

RTPL emission is ascribed to the intrinsic defects in the F:SnO$_2$ film structure. Fig. 4 gives the RTPL spectra of F:SnO$_2$ films at 0, 5, 10, 15 wt
% fluorine level. It is evident that RTPL characteristics of SnO$_2$ films will change with the increased fluorine amount. There are 4 emission peaks observed for pure and 5 wt% F doped tin oxide films, 5 emission peaks for 10, 15 wt% F doped SnO$_2$ films respectively.

Oxygen vacancies (Vo) are the major defects which form the donor levels near the conduction band edge [22]. The NBE UV emission observed at 3.38 eV and 3.34 eVfor 10 and 15 wt% fluorine amount corresponds to the recombination of electrons in the donor levels near the conduction band and holes gathered at the top of the valence band [23,24]. Since the energy of all the emission peaks was lesser than the band gap (3.43–3.98 eV) of all the films, the direct recombination of an electron in the Sn 5p and a hole in O 2p is not observed [25]. Enhancement in the UV emission intensity signifies the increased oxygen vacancy defect on fluorine incorporation. Violet emission peaks observed at 3.09 eV, 3.12 eV, 3.11 eV and 3 eV for all the SnO$_2$ films (0–15 wt% F conc.) attributed to point defects such as oxygen vacancies [26]. Blue emission peaks at 2.59 eV, 2.62 eV, 2.61 eV and 2.60 eV were observed for all the films due to structural defects formed during the film growth [27]. Green emission by all the films at 2.20 eV, 2.23 eV, 2.21 eV and 2.22 eV were due to defect levels formed by singly charged oxygen vacancies [28,29]. A relatively weak red emission was observed at 1.84 eV, 1.85 eV, 1.85 eV and 1.82 eV for 0, 5, 10 and 15 wt% fluorine concentration respectively due to the formation of radiative centers by tin (Sn$_i$) interstitials in the F: SnO$_2$ films [27]. The observed PL emission peaks and the proposed origin of defects were tabulated in Table 4.

Table 2
Root mean square roughness of F:SnO$_2$ thin films.

| F (wt%) | Surface Roughness (nm) |
| --- | --- |
| 0 | 20.5 |
| 5 | 16.6 |
| 10 | 11.3 |
| 15 | 11.2 |

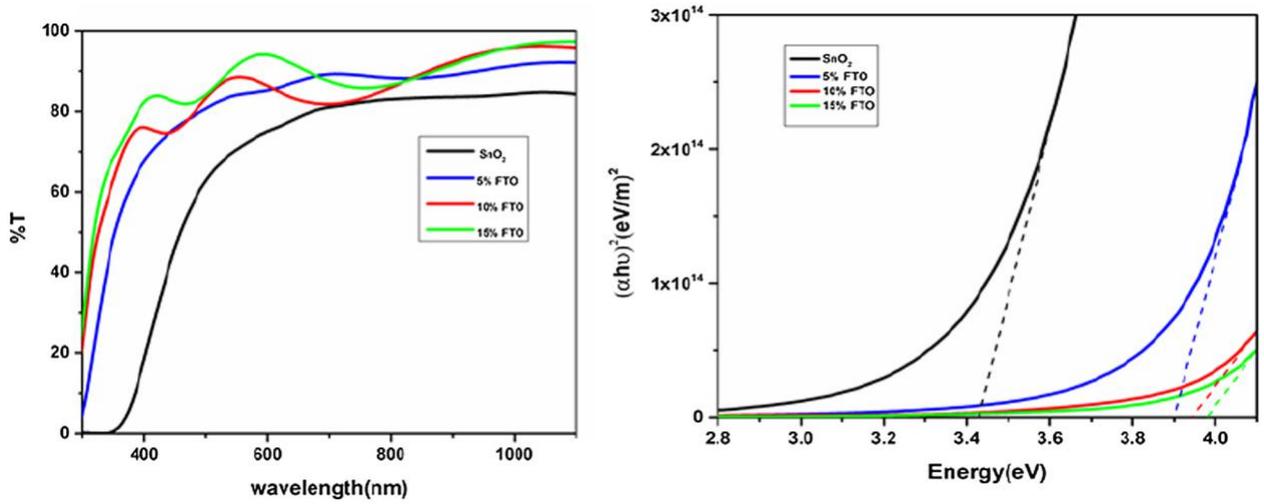

Fig. 3. (a) The transmission spectra and (b) Tauc Plot of F:SnO$_2$ thin films.

Table 3
Band gap of F:SnO$_2$ thin films.

| F conc. (wt%) | 0% | 5% | 10% | 15% |
|---|---|---|---|---|
| Bandgap (eV) | 3.43 | 3.90 | 3.94 | 3.98 |

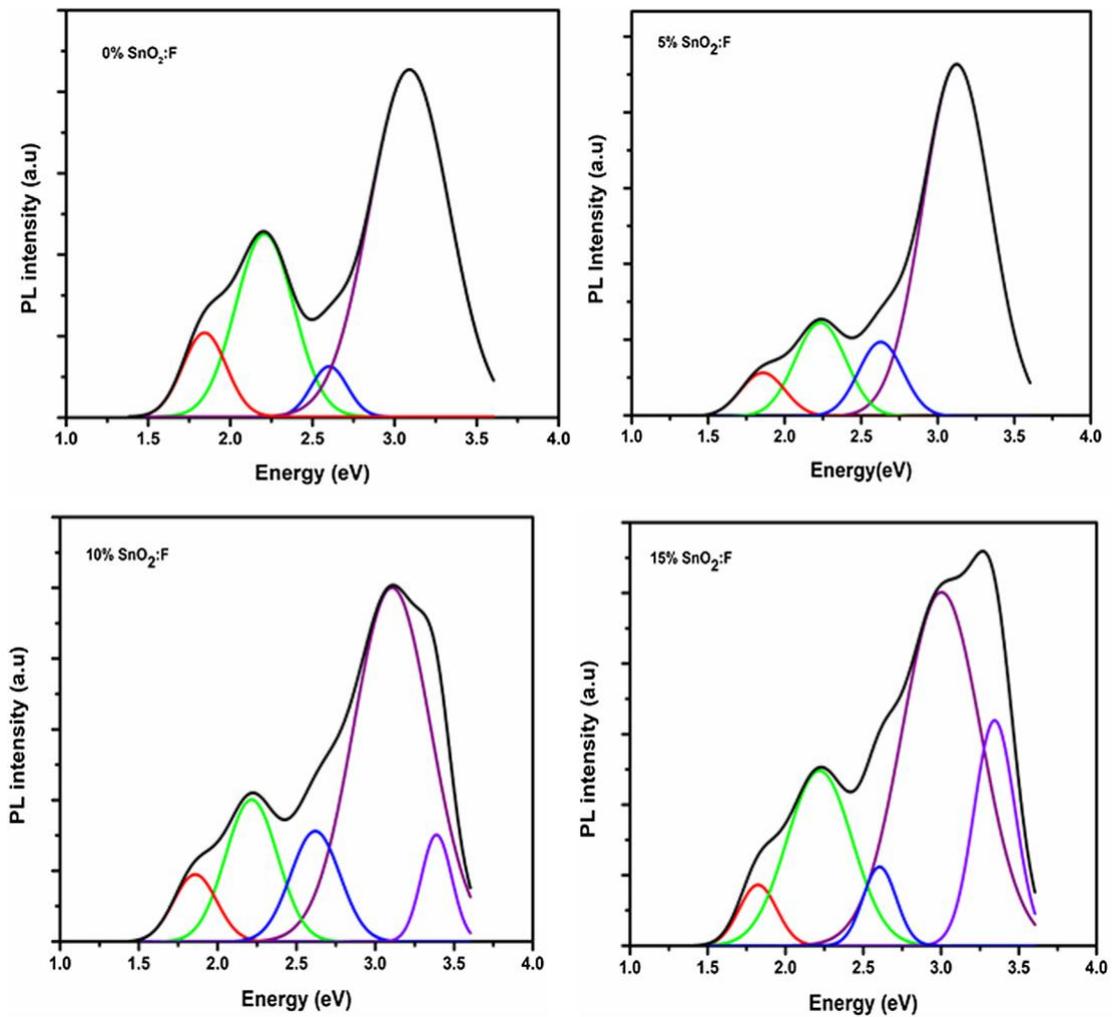

Fig. 4. Deconvoluted photoluminescence spectra of F:SnO$_2$ films at various dopant concentrations.

Table 4
PL emission peaks and proposed origin of defects in F:SnO$_2$ films.

| Undoped | | 5% doped | | 10% doped | | 15% doped | |
| --- | --- | --- | --- | --- | --- | --- | --- |
| Energy (eV) | Defects | Energy (eV) | Defects | Energy (eV) | Defects | Energy (eV) | Defects |
| 1.84 | Sn$_i$ | 1.85 | Sn$_i$ | 1.85 | Sn$_i$ | 1.82 | Sn$_i$ |
| 2.20 | VO+ | 2.23 | VO+ | 2.21 | VO+ | 2.22 | VO+ |
| 2.59 | Structural defects | 2.62 | Structural defects | 2.61 | Structural defects | 2.60 | Structural defects |
| 3.09 | VO | 3.12 | VO | 3.11 | VO | 3.00 | VO |
| | | | | 3.38 | NBE | 3.34 | NBE |

### 3.5. Nonlinear optical studies

#### 3.5.1. Z-scan measurements

Fig. 5 gives the open aperture Z-scan trace of the films which accounts for the nonlinear absorption in the samples. For all the samples, a decrease in the transmittance at the focus was observed which in-dicates that films exhibit reverse saturable absorption (RSA) mechanism [30]. RSA, known as positive nonlinear absorption which arises due to induced scattering mechanisms, two-photon absorption (TPA), free carrier absorption (FCA), excited state absorption (ESA), or combina-tion of all these processes denoted by $\beta_{eff}$ [31]. The condition at which the TPA process occurs in F: SnO$_2$ films was when the energy of the incident light lesser than the energy band gap ($E_g$) of the material but greater than $E_g/2$. In semiconductors, if the incident intensity is large compared to the saturation intensity then the density of states will in-crease near the states that are involved in excitation. Therefore, electrons experience absorption and get excited to the higher energy states before reaching the ground state [32]. Such type of process is called excited state absorption (ESA). When excited state absorption is greater than the ground state absorption, results in the RSA process. In the present case, the source of nonlinear absorption exhibited by the films can be attributed to thermal nonlinearity since the excitation source used were a continuous laser which in turn stimulates only the thermal nonlinear process in the films. Hence, TPA and ESA assisted RSA were the possible mechanism responsible for the nonlinearity in the F: SnO$_2$ films. The obtained data from the Z-scan measurements were fitted using nonlinear transmission equations proposed by Sheikh bahae et al. [30,33]. Nonlinear absorption coefficient $\beta_{eff}$ of the F: SnO2 films are given in Table 5. The obtained values of the nonlinear absorption coefficient ($\beta_{eff}$) were compared with recently reported studies and given in Table 6.

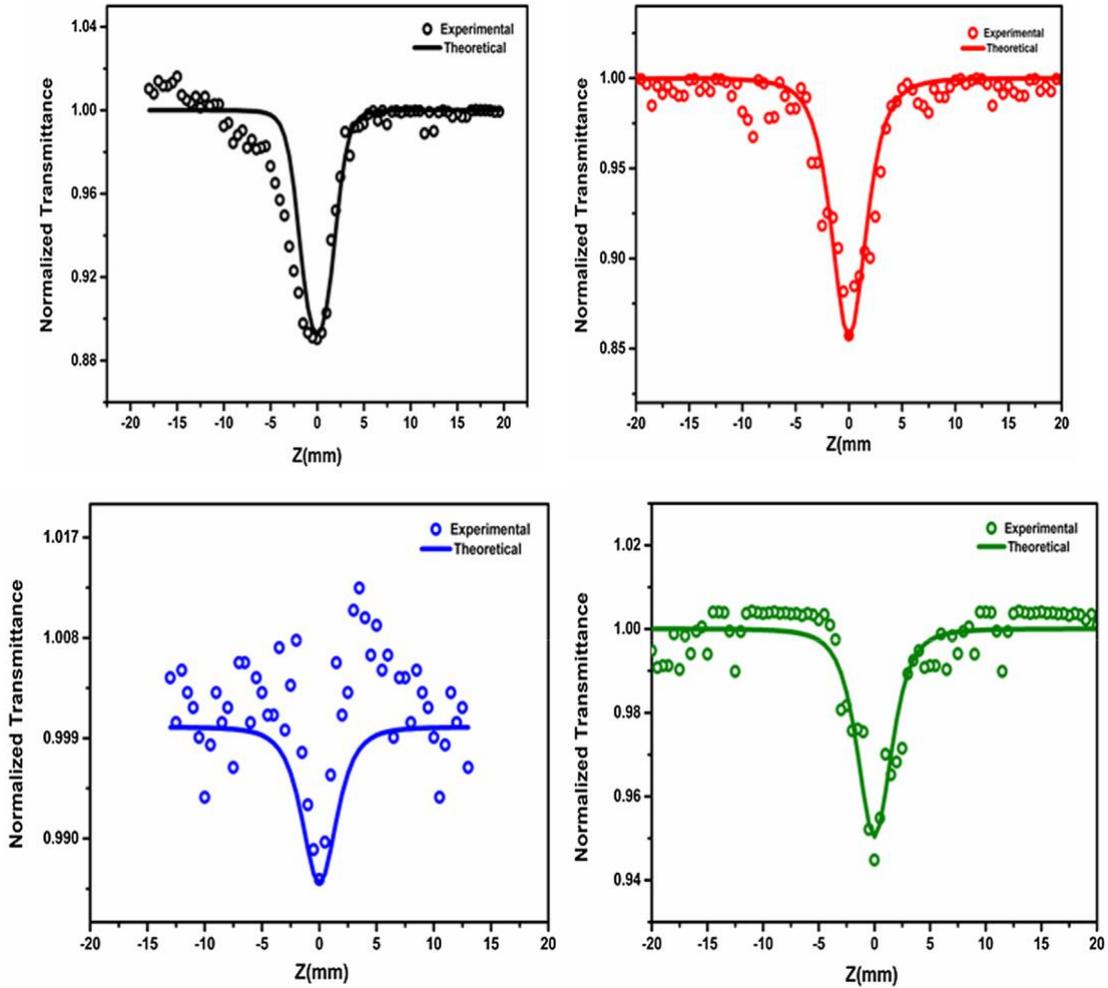

Fig. 5. Open aperture Z-scan traces of F:SnO$_2$ films.

Table 5
Nonlinear absorption coefficient of F:SnO$_2$ thin films.

| F conc. (wt%) | $\beta_{eff} \times 10^{-2}$ cm/W |
|---|---|
| 0 | 9.99 |
| 5 | 13.0 |
| 10 | 1.25 |
| 15 | 5.03 |

Table 6
Recently reported values of nonlinear absorption coefficient.

| Material | Laser regime | $\beta_{eff} \times 10^{-2}$ cm/W |
|---|---|---|
| F: ZnO [30] | He-Ne CW | 1.05–23.30 |
| Gd: ZnO [34] | He-Ne CW | 2.65–8.69 |
| Zn: SnO2 [9] | He-Ne CW | 1.46–9.97 |
| Ba:SnO2 [38] | He-Ne CW | 1.42–3.25 |

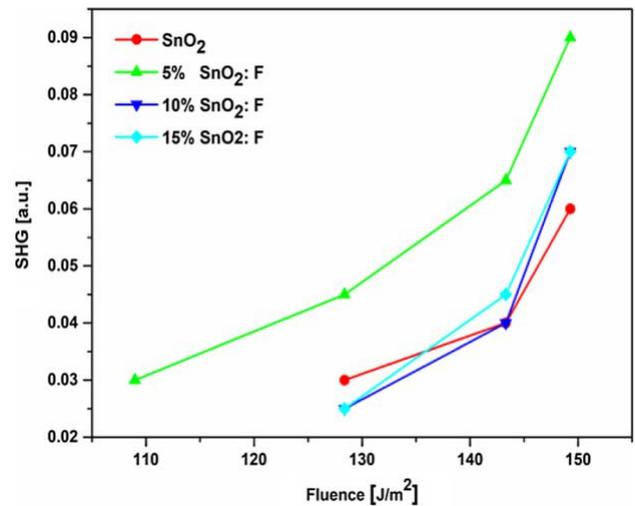

Fig. 6. Laser stimulated SHG.

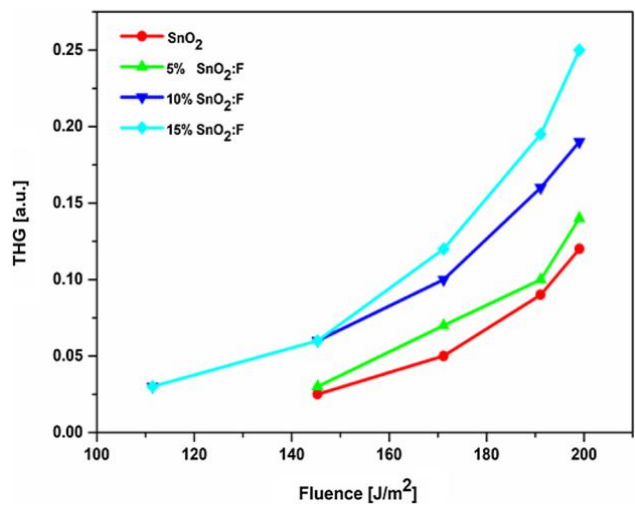

Fig. 7. Laser stimulated.

3.6. Laser stimulated second and third harmonic generation

The measurements of the laser stimulated second and third harmonic (SHG and THG) generation have been performed similarly to the described in the Ref. [34]. The samples are stable with respect to the laser beam power density up to 0.7 GW/cm$^2$. The instability did not exceed 10% and the photo-thermal variations did not exceed 5–10 K. At the beginning the samples have been treated up to 2 min in order to stimulate the nonlinear process, and after they have been illuminated by the probing fundamental laser beam. The studied films presented the mixture of different structural fragments (both acentric as well as centrosymmetric). Despite the SHG is not presented in the centrosymmetric fragments, however, in the laser-induced phenomena induced by coherent laser beams laser nonlinear optical effects described by third-order polar tensors (like SHG) have been occurred. Finally, we deal with the superposition of the NLO effects from different structural clusters. The previous works deal mainly the fixed nonlinear optical constants. In the present work we carry outthe preliminary coherent laser treatment by coherent nanosecond 1064 nm laser pulses during few minutes by coherent light (up to laser stimulated absorption saturation) and after the stabilization, we observed the resulting non-linear optical constants. It is a new origin of the well-known NLO effects which may be used for the development of principally new nonlinear optical triggers and modulators

In order to stabilize SHG and THG signals, we have done the direct measurements with a 20 s interval. To explore the homogeneity of the NLO response with respect to the samples surfaces we have performed the surface scanning by the fundamental 1064 ns laser beams over more than 30- points with the following statistical treatment. The deviations with respect to the averaged magnitude did not exceed 4%. The prin-cipal measurements points are presented in Figs. 6 and 7. Following the presented results one can clearly see that maximal SHG signal was achieved at 5% of the F content and for the THG the maximum is achieved at 15% of the F. Such discrepancies were explained by the different mechanisms of the second and third order susceptibilities. The independent control of the laser stimulated nanosecond coherent treatment for SHG/THG using a thermocouple has shown a fact that the temperature did not exceed 5−10 K which should not be so crucial for the NLO changes. The principal origin SHG for this type of nanostructures is played by the charge density eccentricity due to the F doping. This leads to the occurrence of the flattering sub-bands with the enhanced dipole moments [35,36]. The THG signal intensity shown Fig. 7 shows an enhancement upon fluorine incorporation. The occur-rence of such enhancement may be a consequence of a competition between the photoexcited and relaxation processes on the localized trapping levels [37]. Furthermore, an additional factor which may cause the observed difference here where the occurrence of the multi-photon excitations [37].

4. Conclusions

Fluorine doped Tin Oxide (F:SnO$_2$) films were grown. XRD results showed that deposited films possess polycrystalline orthorhombic structure and also increase in the crystallite size (12.42–32.39 nm) of SnO$_2$ films on fluorine doping. AFM measurements infer that 15%

F:SnO$_2$ film exhibits a very smooth surface. The band gap of the films enhanced from 3.43 to 3.98 eV as doping concentration increased. Intrinsic defects of the films were studied using photoluminescence spectroscopy. Maximal SHG signal was achieved at 5% of the F content and for the THG the maximum is achieved at 15% F:SnO$_2$. Such dis-crepancies were explained by different mechanisms of the second and third order optical susceptibilities. Open aperture Z-scan measurement gives the nonlinear absorption coefficient (β$_{eff}$) of the deposited F:SnO$_2$ films found to be in the order of $10^{-2}$ cm/W. Enhancement in the nonlinear absorption of F:SnO$_2$ films accounts in optical device appli-cations such as optical limiters, frequency converters etc. The observed laser-stimulated NLO effects are principally different with respect to traditional nonlinear optical effects which deal with an interaction of the fundamental laser beam with the material possessing fixed non-linear optical constants. Here we deal with the mentioned effects after their treatment.

Acknowledgements

The presented results are part of a project that has received funding from the European Union's Horizon 2020 research and innovation program under the Marie Skłodowska-Curie grant agreement No 778156. I.V.K., A.W., J.J., acknowledge support from resources for science in the years 2018–2022 granted for the realization of interna-tional co-financed project Nr W13/H2020/2018 (Dec. MNiSW 3871/ H2020/2018/2).